# Inherent structure analysis reveals origin of breakdown of Stokes-Einstein relation in aqueous binary mixtures

Shubham Kumar, Sarmistha Sarkar and Biman Bagchi*

**AFFILIATIONS**

*Solid State and Structural Chemistry Unit, Indian Institute of Science, Bangalore 560012, India.*

*corresponding author's email: bbagchi@iisc.ac.in and profbiman@gmail.com

## Abstract

*We show by inherent structure (IS) analysis that the sharp composition dependent breakdown of the Stokes-Einstein relation correlates surprisingly well with an equally sharp non-monotonic variation in the average inherent structure (IS) energy of these mixtures. Further IS analysis reveals the existence of a unique ground state, stabilized by the optimum number of H-bonds at this composition. The surprisingly sharp turnaround behaviour observed in the effective hydrodynamic radius can be traced back to the formation of low energy equilibrium structures at specific compositions.*

## I. INTRODUCTION

Binary mixtures are long known to exhibit many fascinating anomalies and constitute an important class of solvents. Despite this importance, our understanding of the dynamical properties of these mixtures have remained at infancy. The reason is certainly the complexity of intermolecular interactions among the species. Here we develop a quantitative description based on theoretical analyses of computer simulation results.

Binary mixtures have been widely used as model of glass-forming liquids and the glass transition,[1–4] the potentials employed are simple Lennard-Jones type, with a strong attractive interaction between the two constituents A and B.[5] In the opposite limit of mutually repulsive interaction between the dissimilar species, one can have a scenario where phase separation is narrowly avoided, but a strong composition-dependent anomaly is observed.[6] However, these spherical Lennard-Jones models are too simple to capture the complex structure and dynamics of real, molecular liquid mixtures like carbon tetrachloride-ethyl acetate, methanol-toluene, water-DMSO, water-urea and water-alcohols to name a few. Real mixtures often contain simultaneously both hydrophobic and hydrophilic interactions which make them distinct. This is particularly true for aqueous binary mixtures (ABMs).[7–9]

Aqueous binary mixtures are widely used in chemical, physical and pharmaceutical industries for the reason that these mixtures have highly tunable properties. Even a small, 5-10% change in composition can change the solvation properties greatly.[10–17] ABMs exhibit many remarkable anomalies that have drawn interests over the years.[18–26,27–31] Among aqueous binary mixtures, the strongly non-ideal mixtures of water-DMSO and water-ethanol have received considerable attention not only because of the unique properties of its co-solvents but also due to its various applications in chemical and biological processes. Aqueous DMSO solutions possess anti-inflammatory properties and are used as cryoprotective and pharmaceutical agents[32] whereas aqueous ethanol solutions are used as disinfectant and also in the preparation of tinctures.[33]

Several experimental and simulation studies have identified and reported the composition dependent non-ideality in transport and physicochemical properties of water-DMSO and water-ethanol binary mixtures.[22,34,35] In the case of water-DMSO mixture, Luzar and Chandler[36] rendered correct and almost exact predictions based on simplified mean-field considerations without invoking the concept of significant structures. Later on, many attempts were made to understand the structural origin of the observed anomalies in these mixtures.[37,38] It was proposed that the observed anomalies are due to the formation of 1Solute-nWater (where n = 1 to 4 for both aqueous DMSO and aqueous ethanol mixtures) aggregates. For aqueous DMSO mixtures, Borin and Skaf[39] have found that these complexes coexist in the mixture and their proportion changes with change in the composition of DMSO. However, it always remained unclear how these 1Solute-nWater aggregates can be responsible for the pronounced anomalies observed in these mixtures. Therefore it does not seem to exist an in-depth analysis of the origin of the intriguing properties of these interesting systems.

In his classic treatise entitled "Kinetic Theory of Liquids", J Frenkel[40] envisaged the existence of pre-freezing fluctuations where quasi-stable crystallites form and disappear, much like in the nucleation picture of solids from supercooled melt. Frenkel proposed this as the reason for anomalous rise in viscosity near the freezing transition, or even above and below. This is an appealing picture which has neither been followed experimentally, or simulations, partly because the rise in viscosity near the freezing transition is not very large, certainly not as large as that near the glass transition, and additionally, it is difficult to experimentally detect these crystallites.

One important theoretical technique scantly used in the study of aqueous binary mixtures and one that could identify the quasi-stable fleeting structures in the liquid state is the method of inherent structure determination. Introduced by Stillinger and Weber[1] in the early nineteen eighties, this method has proven to be of great value in identifying the "parent" structures that lie hidden below as low energy structures, but that could influence structure and dynamics at low to intermediate temperature liquids. These inherent structures can be identified or captured by using a technique that combines molecular dynamics simulations with conjugate gradient (CG) technique[1–3] to obtain local minima in energy. In the case of aqueous binary mixtures there could be a multitude of such inherent structures. One important condition would be the conservation of the hydrogen bonds. Another condition would be the balance of hydrophobic and hydrophilic interactions. This is the first ever study to explore the inherent structures in order to explain the anomalous properties of aqueous binary mixtures.

In our recent work,[41] we have explored various aspects including the viscoelastic properties of water-DMSO binary mixtures. We have reported marked deviations from the ideal behaviour in many transport properties such as viscosity, self-diffusion coefficient and orientational relaxation time. The observed non-monotonic behaviour in transport properties is attributed to the local structure of the mixtures and an explanation has been provided in terms of viscoelastic relaxation time. The results are combined with mode coupling analysis which provides additional insights unto the microscopic dynamics of the mixture.

In the continuation of our earlier work,[41] here we have observed several interesting aspects of aqueous binary mixtures, which remained unexplored. We find a sharp turnaround behaviour in the effective hydrodynamic radius of both solute and solvent of water-DMSO and water-ethanol binary mixtures. Moreover, for these binary mixtures, the hydrodynamic radius reveal the existence of a specific composition ($x_C$) where a complete breakdown of the Stokes-Einstein relation is observed. We propose that this could be a universal behaviour in aqueous binary mixtures with amphiphilic solutes. We have shown through inherent structure analysis that at a critical composition ($x_C$), the average inherent structure energy exhibits a sharp fall, followed by a rise, for both water-DMSO and water-ethanol binary mixtures. We have also provided explanations in terms of the formation of an open network quasi-stable structures which is stabilized by the optimum number of hydrogen bonds between water and solute (DMSO/EtOH) at the composition corresponding to the turnaround point.

The organization of the rest of the paper is as follows. In **Section II**, we discuss the potential models and the technical details employed in the simulations. In **Section III**, we discuss the theoretical background addressing the methodologies employed to compute different properties of aqueous binary mixtures. The detailed numerical results obtained from our simulations are presented in **Section IV**. In this result-section, at first, we discuss the turnaround behaviour of the effective hydrodynamic radius when plotted against viscosity. Subsequently, we have performed the inherent structure analysis to explore reasons behind the breakdown of Stokes-Einstein relation and also the anomalous properties of ABMs. Furthermore, we discuss the microscopic origin of the observed turnaround behaviour that is due to the formation of open networked transient structures. In **Section V**, we present a concise discussion of the results along with concluding remarks in connection with future problems.

## II. SIMULATION DETAILS

Molecular dynamics simulations of water-DMSO and water-ethanol binary mixtures having a total of 1000 molecules corresponding to various mole fractions of 0.0, 0.05, 0.10, 0.15, 0.20, 0.25, 0.30, 0.35, 0.40, 0.50, 0.60, 0.70, 0.80, 0.90 and 1.0 are carried out using GROMACS package (version 5.0.7).[42] For water, we have used two different models namely a three-site SPC/E model[43] and a four-site TIP4P/2005 model[44] while for DMSO and ethanol, we have used the four site flexible GROMOS96 53a6 potential of Oostenbrink *et. al.* [45]
We have performed energy minimization using steepest descent algorithm followed by equilibration runs in isothermal-isobaric (NPT; P=1bar and T=300 K) and canonical (NVT; T=300 K) ensembles respectively for 10 ns each. For the calculation of viscosity, diffusion coefficient and number of hydrogen bonds we have carried out the production run of 15 ns in canonical ensemble (NVT; T=300 K) *using* Nosé–Hoover thermostat.[46] The equations of motion have been integrated using leapfrog algorithm[47] with a time step of 1 fs. The coordinates, components of velocities and forces have been saved every 2 fs. The cut-off radius was set to $r_c = 1.2\,nm$ for LJ interactions. Electrostatic long-range corrections have been considered by the particle mesh Ewald (PME) method.[48] All the simulations have been carried out in cubic boxes with periodic boundary conditions. To calculate the inherent structure energies, we have performed the conjugate gradient (CG) minimization[49] for 2000 equilibrium configurations and obtained the average inherent structure energies.

## III. FUNDAMENTALS

### A. Calculation of viscosity

The shear viscosity (η) of binary mixtures is calculated using Green-Kubo relation of shear viscosity as given below,

$$\eta = \frac{1}{Vk_B T} \int_0^\infty dt \left\langle \sigma^{\alpha\beta}(t)\, \sigma^{\alpha\beta}(0) \right\rangle \qquad (1)$$

where $V$ is the volume of the system, $k_B$ is the Boltzmann constant, T is the temperature, $\sigma^{\alpha\beta}$ is the off-diagonal element of the stress tensor, defined as

$$\sigma^{\alpha\beta} = \sum_{i=1}^{N} \left( m_i v_i^\alpha v_i^\beta + \frac{1}{2} \sum_{\substack{j=1 \\ j\neq i}}^{N} F_{ij}^\beta r_{ij}^\alpha \right) \qquad (2)$$

where $m_i$ is the mass, $v_i^\alpha$ and $v_i^\beta$ are the $\alpha$ and $\beta$ components of velocity of the $i^{th}$ particle; $F_{ij}^\beta$ is the $\beta$ component of the force exerted on particle $i$ by particle $j$ and $r_{ij}^\alpha$ is the $\alpha$ component of the particle-particle separation vector, $r_{ij} \equiv r_j - r_i$. The indices $\alpha, \beta = x, y, z$ and $\alpha \neq \beta$. In our calculations of shear viscosity, we have taken the average of the six stress-autocorrelation functions obtained by the six components of the off-diagonal stress tensor, i.e. $\sigma^{xy}$, $\sigma^{yx}$, $\sigma^{xz}$, $\sigma^{zx}$, $\sigma^{yz}$ and $\sigma^{zy}$, that enhances the accuracy of the results.

### B. Calculation of diffusion coefficient

The self-diffusion coefficient ($D_i$) of the components of binary mixtures is calculated using the Green-Kubo relation of shear viscosity as given below,

$$D_i = \frac{1}{3N_i} \int_0^\infty dt \left\langle v_k(t)\, v_k(0) \right\rangle \qquad (3)$$

where $v_k(t)$ is the velocity of the centre of mass of molecule $k$ at time $t$ and $N_i$ is the number of molecules of type $i$.

### C. Computational details of inherent structure analysis

By removing the thermal motions completely from a specific configuration of the liquid, the new structure belonging to a local potential energy minimum is obtained. Such a structure is commonly known as the inherent structure of that liquid. In computer simulations, inherent structures are obtained by quenching the equilibrium configurations to the corresponding local minima using the conjugate gradient (CG) algorithm.[1,3,49] We carry out the CG minimizations for 2000 equilibrium configurations from which we have computed the inherent structure energies.

**D. Criteria for the calculation of the number of hydrogen bonds**

We have applied a similar geometric criterion as described by Kelin *et. al.*[50] For hydrogen bonding, the molecules need to fulfil the following three geometrical criteria:

(a) The distance RO-O between the donor and the acceptor should be shorter than the threshold value 3.56 Å.

(b) The distance RO-H between the donor and the acceptor should be shorter than the threshold value 2.48 Å.

(c) The angle O-H --- O should be smaller than the threshold value of $30^0$.

## IV. RESULTS AND DISCUSSION

**A. Turnaround (re-entrant) behaviour of the effective hydrodynamic radius with viscosity: Breakdown of Stokes-Einstein relation**

The conventional Stokes-Einstein (SE) formula is given by[51,52]

$$D = Ck_B T/a\eta \qquad (5)$$

where $D$ is diffusion coefficient, $\eta$ is shear viscosity, $a$ is molecular radius, $k_B$ is the Boltzmann constant, $T$ is the temperature and $C$ is a numerical constant.

According to conventional SE relation, $1/D$ varies linearly with $\eta$. However, deviations from Stokes-Einstein relation has been observed in many cases. In the seminal work done by Zwanzig and Harrison,[53] it was suggested that the deviation from conventional SE relation can be explained in terms of an effective hydrodynamic radius (EHR), given by the expression

$$C/R_H = \eta D/k_B T \qquad (6)$$

where $R_H$ is the effective hydrodynamic radius. **Figure 1** depicts the variation of the inverse of the effective hydrodynamic radius (EHR) of solute and solvent molecules as a function of viscosity of water-DMSO and water-ethanol mixtures.

With both the water models employed in the present study, we find that the EHR of both solute and solvent shows the turnaround (re-entrant) behaviour, when plotted against viscosity, for water-DMSO and water-ethanol binary mixtures at $x_{DMSO}$ ~ 0.35 and $x_{EtOH}$ ~ 0.25 respectively. The effective hydrodynamic radius (EHR) of a molecule captures the essence of interaction of the surrounding molecules. This sharp turnaround behaviour in the effective hydrodynamic radius reveals that the effective interaction between the solute and the solvent, and also among themselves undergo a dramatic change at a specific composition.

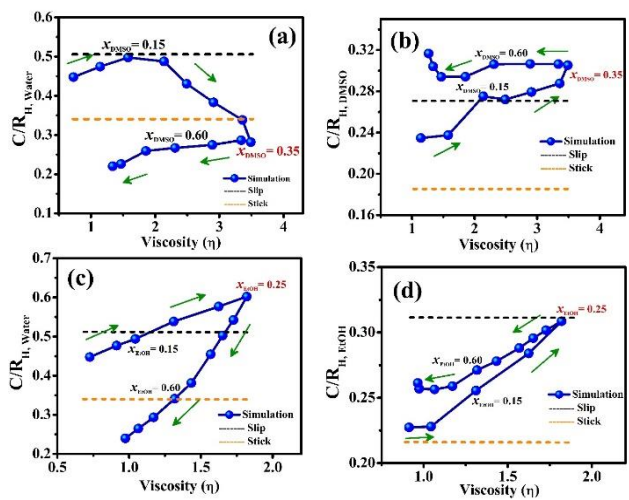

**Figure 1. The inverse of the effective hydrodynamic radius (C/R$_H$) showing turnaround (re-entrant) behaviour for (a) SPC/E water and (b) DMSO in water-DMSO mixtures; (c) SPC/E water and (d) ethanol in water-ethanol mixtures respectively when plotted against simulated viscosity for these mixtures. The value of R is given in the unit of nm. C is $1/4\pi$ for slip boundary conditions and $1/6\pi$ for stick boundary conditions. Note that the orange and black dotted lines represent the ideal value of the inverse of the molecular radius as predicted by stick and slip boundary conditions, respectively. The direction of the green arrows shows the increasing concentration of solute.**

This is of course not new, over the years the effective hydrodynamic radius has been used as a measure of effective solute-solvent interactions. As envisaged in hydrodynamics, this radius provides a measure of the coupling of the tagged solute with the surrounding solvent.

In **Figure 1,** we also depict the values predicted by hydrodynamics, both for the slip and the stick boundary conditions,[51,54–56] with molecular radius. The hydrodynamic values provide a valuable reference for the present discussion. Although hydrodynamics completely fails to reproduce the turnaround or re-entrant behaviour, the predicted values are not too away from the observed values. How do we understand such anomalous turnaround? The dynamic properties such as viscosity and diffusion coefficient of binary mixtures derive a huge contribution from the local structures. However, diffusion and viscosity usually obey their inverse relationship. We find that the anomaly observed beyond a particular composition is a manifestation of the existence of a unique ground state at that particular composition. In the case of water-DMSO mixture, that particular composition is around $x_{DMSO} \sim 0.35$ whereas for water-ethanol mixture, it is around $x_{EtOH} \sim 0.25$. In order to quantify the unique ground state, we have explored the inherent structure analysis which is successful in explaining the origin of anomalous properties of ABMs. In the forthcoming section, we have described the details of the inherent structure analysis.

## B. Inherent structure analysis in explaining the turn around behaviour and the breakdown of stokes-Einstein relation

Inherent structure analysis is capable of providing an insight into the microscopic structural details of the parent liquid.[3] One can obtain the inherent structures by removing the kinetic energy, vibrations of the molecules of the mixtures. In **Figure 2(a),** we have plotted the average inherent structure energies of water-DMSO as a function of mole fraction of DMSO for both the water models (SPC/E and TIP4P/2005), whereas in **Figure 2(b)**, we have shown the average IS energies of water-ethanol mixtures as a function of mole fraction of ethanol.

We have found that the anomaly observed at the particular composition is a manifestation of the existence of a unique ground state. In case of water-DMSO mixture, that particular composition is around $x_{DMSO} \sim 0.35$ whereas for water-ethanol mixture, it is around $x_{EtOH} \sim 0.25$. This is clearly reflected in the plots of inherent structure energies as a function of composition for both water-DMSO and water-ethanol binary mixtures where the minima are found at $x_{DMSO} \sim 0.35$ and $x_{EtOH} \sim 0.25$ respectively, and as shown in **Figures 2(a) and 2(b).** Hence, the clear signature of the presence of unique ground states are verified by these minima of inherent structure energies.

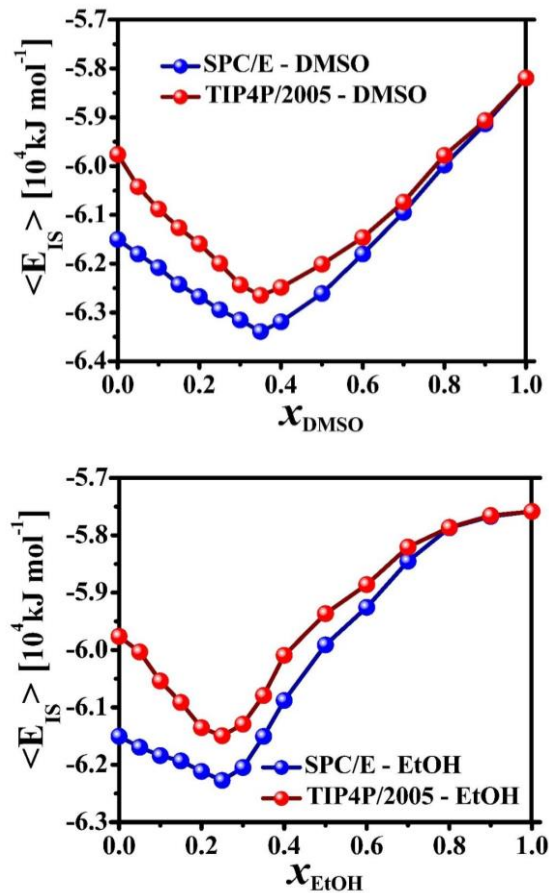

**Figure 2. Plots of inherent structure energies as a function of composition for (a) water-DMSO and (b) water-ethanol binary mixtures. Please note that for water-DMSO mixture it is showing minimum at $x_{DMSO} \sim 0.35$ whereas for water-ethanol mixture, it is around $x_{EtOH} \sim 0.25$. This is the signature of the presence of unique ground states of these two binary mixtures corresponding to the particular compositions.**

## C. Microscopic characterization of unique ground state through partial radial distribution functions of inherent structures

In order to characterize the microscopic details of unique ground state, we have calculated the radial distribution function (RDF) between the centre of mass (COM) of solute (DMSO/EtOH) and solvent (water) of parent liquids and its corresponding inherent structures. **Figure 3(a) and 3(b)** depicts the comparison study of RDF of parent liquids and its corresponding inherent structures for water-DMSO at $x_{DMSO} \sim 0.35$ and water-EtOH mixtures at $x_{EtOH} \sim 0.25$ respectively.

As expected, the peaks of RDF of inherent structures are sharper compared to that of the parent structures for both water-DMSO and water-EtOH mixtures.

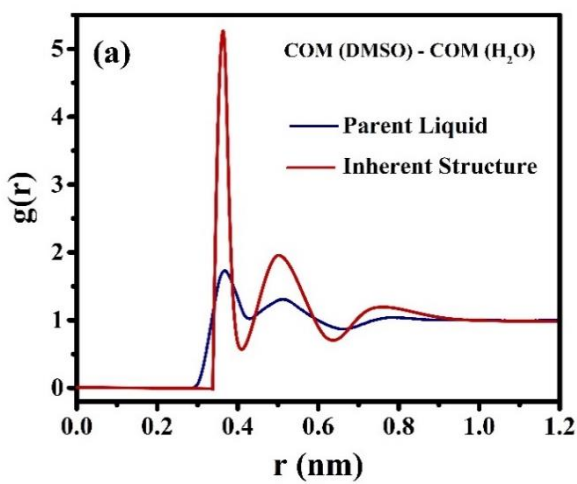

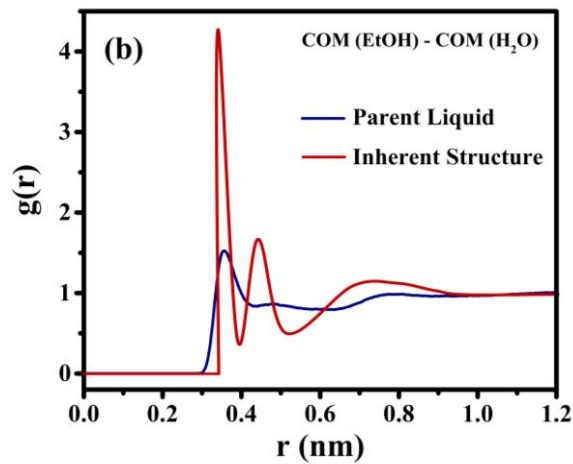

**Figure 3.** Plot of radial distribution function between center of mass (COM) of solute and center of mass (COM) of SPC/E water for parent liquid and its corresponding inherent structure at (a) $x_{DMSO}$=0.35 for water-DMSO and (b) $x_{EtOH}$=0.25 for water-EtOH.

**Figure 4(a) and 4(b)** depict the variation of the first peak height of RDF of inherent structures with respect to the mole fraction of solute for water-DMSO and water–ethanol mixtures respectively. Here also, we have observed the signature of non-monotonic behaviour in mole fraction dependence.

**Figure 5(a) and 5(b)** show the wave number dependent structural relaxation time, $\tau_F(k)$ of SPC/E water-DMSO mixtures against mole fraction of DMSO at $k = 1.5$ Å$^{-1}$ and $k = 2.5$ Å$^{-1}$ respectively. For both the values of $k$, a maxima is found at $x_{DMSO} \sim 0.35$. Surprisingly, both the plots (**Figure 4 and Figure 5**) show same kind of non-

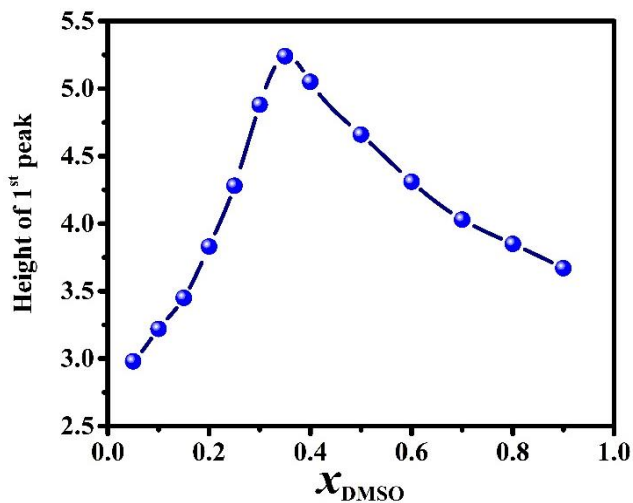

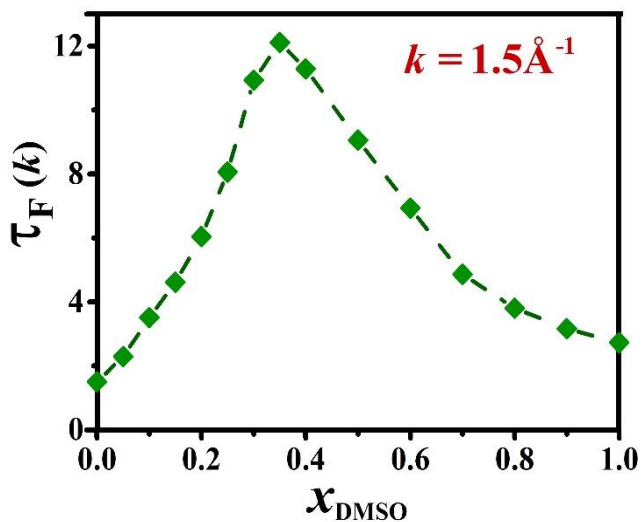

(a)

(a)

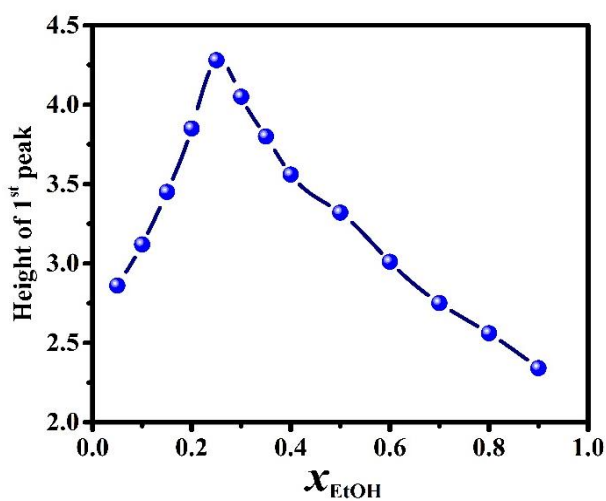

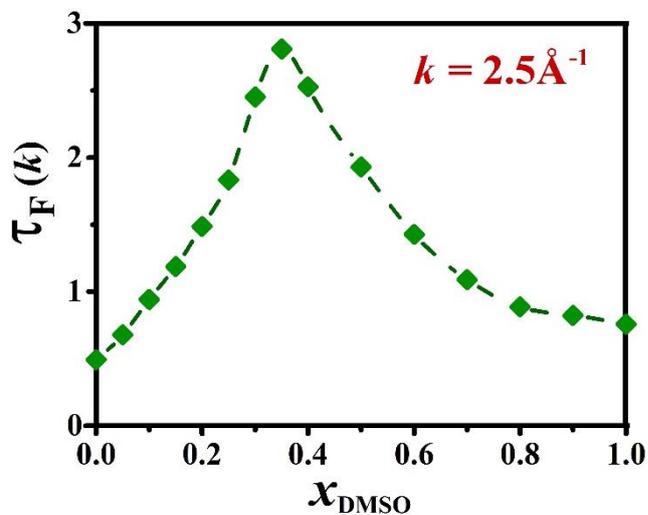

(b)

(b)

**Figure 4.** The heights of the first peak of RDF of inherent structures are plotted against mole fraction of solute for (a) water-DMSO and (d) water–ethanol mixtures. It is to be noted that it shows the similar behavior with the structural relaxation time as shown in the Figure 5.

**Figure 5.** The wave number dependent structural relaxation time, $\tau_F(k)$ of SPC/E water-DMSO mixtures as a function of composition of DMSO at (a) $k = 1.5$ Å$^{-1}$ and (b) $k = 2.5$ Å$^{-1}$. For both the values of $k$, a maxima is obtained at $x_{DMSO} \sim 0.35$.

**Figure 6(a) and 6(b)** show the mole fraction dependence of the coordination numbers of water around the COM of solute (DMSO/EtOH) in the first solvation shell. This is to be noted that the calculations of coordination numbers (as shown in Figure 6(a) and 6(b)) are performed on inherent structures. The discontinuity at $x_{DMSO}$ ~ 0.35 for water-DMSO mixture and at $x_{EtOH}$ ~ 0.25 for water-EtOH mixture signifies the structural transformaion beyond those particular compositions for both the mixtures.

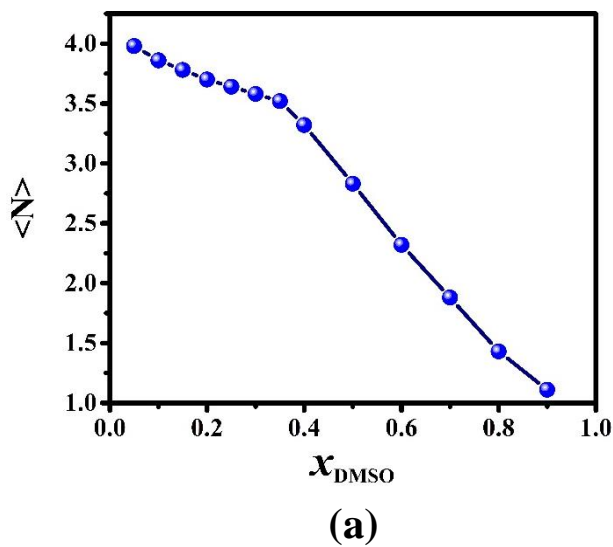

(a)

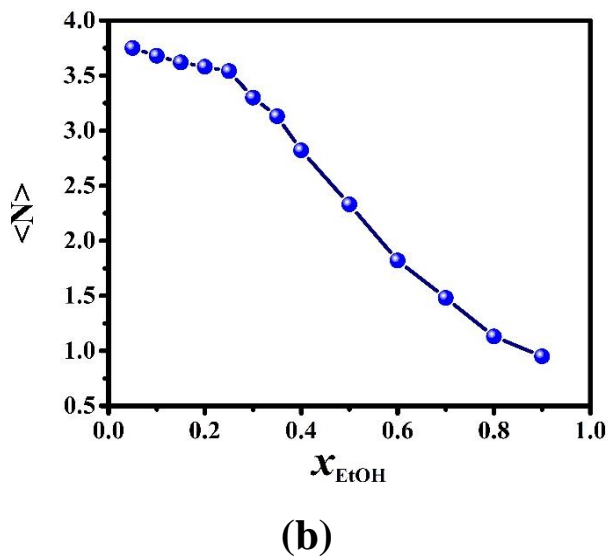

(b)

**Figure 6.** Composition dependence of the coordination number of water around (a) DMSO in water-DMSO mixture and (b) Ethanol in water-ethanol mixture obtained from the inherent structures.

The above results serve to explain the surprising turn around behaviour in the effective hydrodynamic radius described earlier. It is clear that there exists particular composition where the mixture acquires extra stability due to the fulfilment of the hydrogen bonding requirement. The structure thus gains additional stability.

The above results can also be used in a mode coupling theory (MCT)[57,58] description that employs wave number dependent dynamic structure factor (DSF). In MCT, viscosity is obtained by a wavenumber integration of the dynamic structure factor. Thus, non-monotonic dependence in the relaxation of DSF is reflected in the non-monotonic dependence of viscosity, and in effective hydrodynamic radius.

**D. Hydrogen-Bond Analysis**

Another useful order parameter to describe the observed anomalies is the number of water-solute (DMSO/EtOH) hydrogen bonds. Here we show, by computer simulations of water-dimethyl sulfoxide (W-DMSO) and water-ethanol (W-EtOH) mixtures that the anomalies can be traced back to the existence of a unique, quasi-stable, maximally H-bonded network structure, stabilized further by hydrophobic interactions. We calculate the number of hydrogen bonds between water and solute (DMSO/EtOH). **Figure 7(a)** depicts the total number of hydrogen bonds between water and DMSO per time frame as a function of mole fraction of DMSO. In **Figure 7(c)** the total number of hydrogen bonds between water and ethanol as a function of concentration of ethanol is shown. For water-DMSO mixtures, it exhibits a maximum near mole fraction 0.35 of DMSO whereas, for water-ethanol mixtures, it exhibits a maximum near mole fraction 0.25 of ethanol. In **Figure 7(b)**, we show the average number of hydrogen between water and DMSO per time frame per DMSO molecule. The analogous quantity for water-ethanol mixture is shown in **Figure 7(d).** The average number of hydrogen bonds per solute molecule shows an abrupt fall after $x_{DMSO} \sim 0.35$ for water-DMSO mixture whereas in case of water-ethanol mixture the same kind of abrupt fall is observed after $x_{EtOH} \sim 0.25$.

To understand the microscopic origin of the unusual composition dependence of the transport properties, we carried out structural analysis. In the case of water-DMSO mixture, the composition 0.35 of DMSO mole fraction is unique because this is the mole fraction where the hydrogen bond condition of one DMSO with two water molecules is best satisfied.

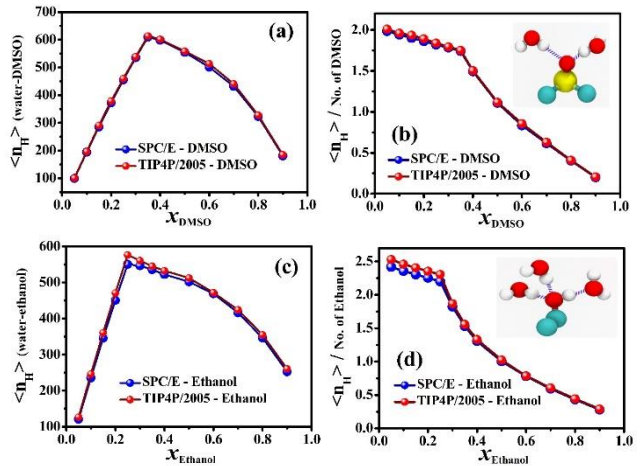

**Figure 7. The total number of water-solute (DMSO/EtOH) hydrogen bonds per timeframe against composition for (a) water- DMSO and (c) water-ethanol solutions. For water-DMSO mixture, a maximum in the total number of solute-solvent hydrogen bonds is observed at $x_{DMSO}$ ~ 0.35, whereas in the case of water-ethanol mixture, it is observed at $x_{EtOH}$ ~ 0.25. The average number of water-solute (DMSO/EtOH) hydrogen bonds per time frame per solute molecule against composition for (b) water-DMSO and (d) water-ethanol solutions. For water-DMSO mixture, a sharp fall in the average number of hydrogen bonds is observed after $x_{DMSO}$ ~ 0.35, whereas in the case of water-ethanol mixture, it is observed after $x_{EtOH}$ ~ 0.25.**

That is, the hydrogen bond network between water and DMSO is least frustrated. Due to the much stronger hydrogen bond between water hydrogen and DMSO oxygen, the hydrogen bond network is not only the most stable but also adopts an open network like structure which is further supported by the hydrophobic interaction among the methyl groups.

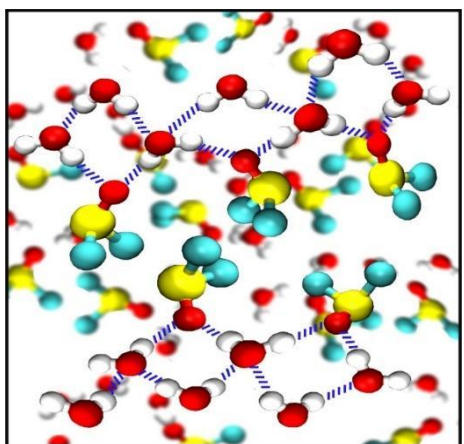

**(a)**

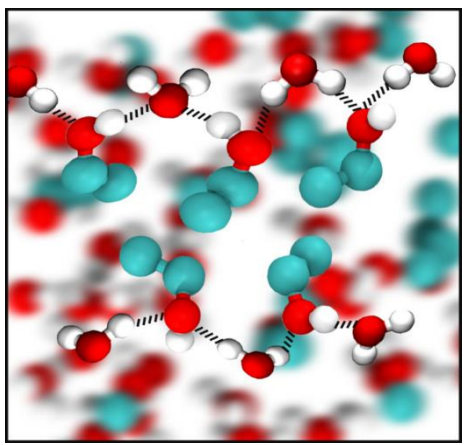

**(b)**

**Figure 8. Snapshots of extended transient structures formed at (a) DMSO concentration of b) ethanol concentration of $x_{EtOH}$ ~ 0.25. In both cases, the network structures are formed through H-bonding as well as alkyl-alkyl aggregation of solute (DMSO/ethanol) molecules. The intermolecular hydrogen bonds in the transient structure are shown by dashed blue lines and dashed black lines in (a) and (b) respectively. The methyl groups (as united atoms) are coloured as cyan, the sulfur atoms as yellow, the oxygen atoms of both water and solute as red and hydrogen atoms of water are represented by white balls.**

In the case of water-ethanol mixtures, the same scenario unfolds but at a lower ethanol composition. Here the most stable arrangement occurs at the ethanol composition 0.25 because ethanol forms three hydrogen bonds with one water molecule. Thus, the principle of minimum frustration of hydrogen bonds suggests the composition ¼ which is indeed observed in simulations. The snapshots open network structures of water-DMSO mixture (at $x_{DMSO}$ =0.35) and water-ethanol mixture (at $x_{EtOH}$ ~ 0.25) are shown in **Figure 8(a) and 8(b)**. These structures, however, are transient having lifetimes of few picoseconds but nevertheless can slow down the dynamics of the system in a remarkable way.

## V. CONCLUSION

The sharp turnaround (or**,** re-entrant) behaviour observed in water-DMSO and water-EtOH binary mixtures is indeed surprising. The microscopic origin of this re-entrance is even more surprising, especially so because the reason is the same for the two mixtures. We established that both the re-entrances are consequence of the existence of a unique stable state, at the respective composition where each mixture is structurally the most stable. We further demonstrated that this stability is due to the appearance of a network with an optimum number of the hydrogen bond, in the sense that the frustration in the number of hydrogen bonds is minimum. In both cases, the hydrogen bond network is further stabilized by hydrophobic interactions (as shown in **Figure 8(a) and 8(b)**).

The sharp re-entrance or turnaround in these binary mixtures is a manifestation of the existence of a unique stable state at a composition which is different for water-DMSO and water-EtOH mixtures. The unique stable state is a consequence of the amphiphilic character of DMSO and ethanol in water. This unique ground state is quantified by the intherent structure analysis. What is unique and surprising is that at the said compositions, both the hydrogen bond criteria and the hydrophobic interactions between the methyl (for DMSO) and ethyl (for ethanol) are globally satisfied to the maximum extent. This gives the aqueous solution at those compositions an uncommon character, reflected in the turnaround of the effective hydrodynamic radius, $R_H$ (**Figure 1**), revealing a total breakdown of hydrodynamic predictions. While Zwanzig and Harrison[53] elegantly pointed out the variation of effective hydrodynamic radius with changing nature of the solvent, we are not aware any previous demonstration of such sharp variation. We expect such re-entrance of effective hydrodynamic radius to be a general feature of the amphiphilic solutes in water, and could actually be used to characterize the nature of these solutions. And finally, a fully theoretical calculation would involve a complex and non-trivial implementation of mode coupling theory type approach[57] which is under progress.

## ACKNOWLEDGEMENTS
BB thanks Department of Science and Technology (DST, India) and Sir J. C. Bose Fellowship for providing partial financial support. S. Kumar thanks Council of Scientific and Industrial Research (CSIR) for the research fellowship. S. Sarkar thanks Department of Science and Technology (DST, India) for her Post-Doctoral fellowship.